The oldest evidence about utilization of Saffron dates back to 2,000-1,500 years BC. Saffron is a triploid sterile species that is vegetatively propagation with Corm that only survives for only one season,reproducing via division into "cormlets" that eventually give rise to new plant and took time at least one year for each plants ,Each saffron flower has three stigmata and one stigma weights about 2mg. It takes 150,000-200,000 flowers and over 400 h of hand labor to produce 1kg saffron stigma.Now a days by studying in RT-PCR,Real Time PCR,tissue culturing,... tecniques with the help of Molecular science and biotechnology researchs we can produce Saffron widley in laboratory any time we need free of contaminants.Here,in this book Mandana Mirbaksh and Dr.Faraz Zandiyeh reviewed on important issues and published their investigation on saffron plants,its morphology,distribution,growth,medical uses,Implication of Carotenoid Biosynthesis Genes during stigma development,...They have also represented their molecular researchs in RNA extraction,cDNA preparation,RT-PCR tecnique and caught significant results.

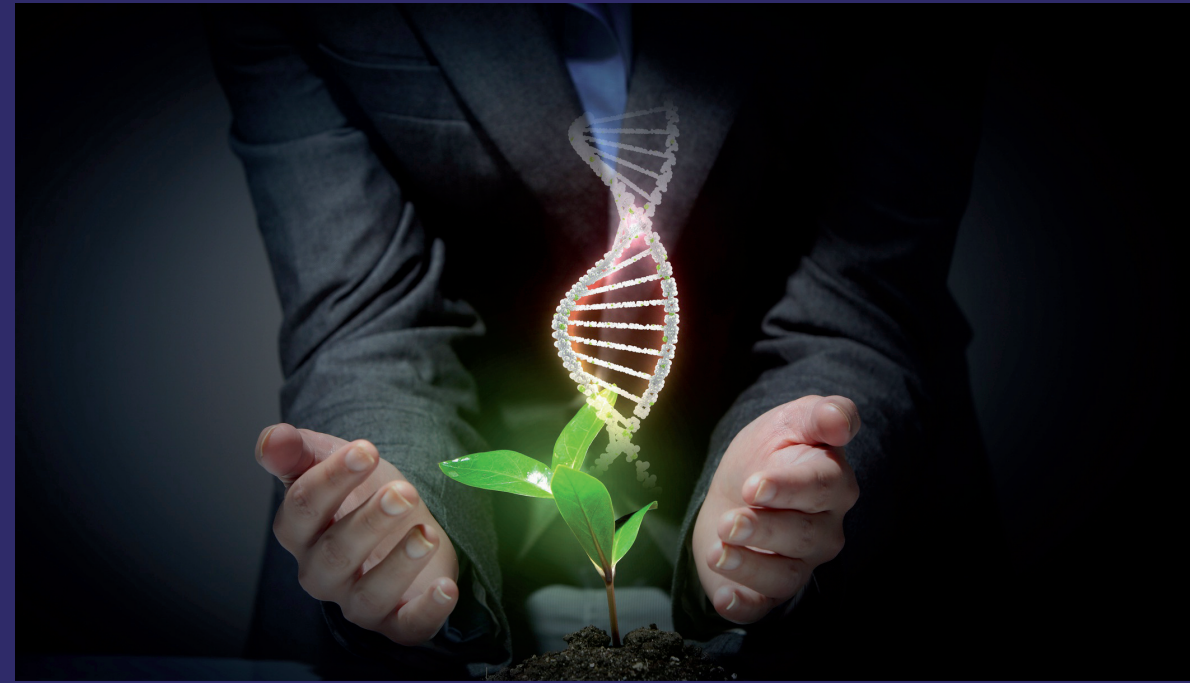
Mandana Mirbakhsh

Faraz Zandiyeh

# In vitro micropropagation and apocarotenoid gene expression in saffron

Investigation of in vitro micropropagation and apocarotenoid gene expression in perianth of saffron (Crocus sativus L.)

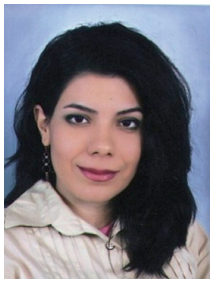

**Mandana Mirbakhsh**

Mandana Mirbakhsh is born on May 1987 in Tehran,Iran;she holds a M.S in plant Physiology from Alzahra University in 2013.This book is the part of her research by the help of Dr.Faraz Zandiyeh who is born on August 1987 in Tehran,Iran;gratuated from shahid Beheshti Medical University in 2013 and wrote 'Population and Family planning' book in 2013.

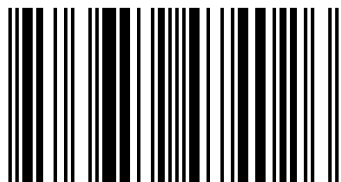

978-3-659-47402-6

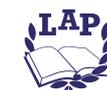

**Mandana Mirbakhsh**
**Faraz Zandiyeh**

**In vitro micropropagation and apocarotenoid gene expression in saffron**

Mandana Mirbakhsh
Faraz Zandiyeh

# In vitro micropropagation and apocarotenoid gene expression in saffron

Investigation of in vitro micropropagation and apocarotenoid gene expression in perianth of saffron (Crocus sativus L.)

# Investigation of *in vitro* micropropagation and apocarotenoid gene expression in perianth of saffron
# (Crocus sativus L.)

# Authors :

Mandana Mirbakhsh (Master of plant physiology Science)

Biology Department, Science Faculty, Alzahra University, Tehran, Iran.

Email: mandana.mirbakhsh.66@gmail.com

Faraz Zandiyeh (M.D)

Gratuated from Shahid Beheshti Medical University ( SMBU ) ,Tehran,Iran.

Email: Faraz.dr@gmail.com

The authors would like to thank Dr. Monir.H.Namin of Herbaceous Physiology department of Alzahra University of Tehran, Iran. The best advisor and teacher that anyone can hope for. None of the achievements in this book could have been made without her support and help. Not only as an advisor professor in this research she lightened our way and thought us so much but also her experience and knowledge in this delicate area made us obtain the results printed in this manuscript.

We would like to thank Mr. Shooshtari, for his assistance in the field work and helping us collecting samples of Saffron flowers from Ghaen district of Khorasan province in Iran. The research would not have been accomplished without those particular and well collected samples.

We would also like to thank everyone whom helps us in one way or other. We appreciate all the support and help we received from all our colleges and friends : Narmin Najafzedeh , Zahra Zahed, and Sepideh Mashyekhi in sample collection, lab work, analyzing the data and molecular investigations.

Mandana Mirbakhsh(M.S)

Faraz Zandiyeh (M.D)

To Amineh SafaeiRad & Nahid Varzandi

Our angels, our hearts, our inspirations;

We owe you our lives.

You are the rock of our lives

You gave us wings to fly

We owe our accomplishments to you.

This manuscript is dedicated to you and all amazing people in our lives whom believed in us and our abilities.

# Abrtact

Saffron (*Crocus sativus L.*) is a triploid, sterile, monocot plant belongs to the family Iridaceae, sub family Crocoideae. C.sativus only bloom once a year and should be collected within a very short duration, the stigmas of Saffron flowers are harvested manually and subjected to desiccation then have been used as a spice. It has been also used as a drug to treat tumor, cancer, chronic uterine hemorrhage, insomnia, scarlet fever, small pox, colds and cardiovascular disorders. It has been shown that saffron is a protective agent against chromosomal damage. Saffron has been vegetative propagated by corm, each mother corm produce 7-8 cormlet each year. The main colors of saffron, crocetin and crocetin glycosides, and the main flavors, picrocrocin, Safranal is the main component of aroma and It's bitter taste is related to Glycoside picrocrocin that are derived from the oxidative cleavage of the carotenoid, zeaxanthin.by zeaxanthin cleavage dioxygenase (ZCD). We investigated gene expression of ZCD in vitro by using tissue culture of perianth obtained from immature flora buds of Ghaen of Khorasan province, Iran, on MS medium supplemented by 10 mg/L NAA and BAP. RNA of each sample was extracted using RNx method; followed by RT-PCR techniques .The results indicated that ZCD was present in perianth of all cultured samples of mentioned areas. Investigation of this pathway which controls the saffron apocarotenoid pigments in prianth is important to produce saffron with high quality and quantity. The results showed that the expression of ZCD in perianth whether through mevalonic acid or non-mevalonic acid needs more investigation.

# Chapter 1

# SAFFRON

# 1. Saffron

## 1.1 Introduction

The dried red stigma of *Crocus sativus* L., that is widely used as a spice, and drug is called Saffron, this valuable crop is cultivated for at least 3,500 years. Some archaeological and historical studies indicate that domestication of saffron date back to 2,000-1,500 years BC [1] ,But The oldest evidence about utilization of Saffron dates back to an ancient Persian dynasty "Achaemenian".
The name Saffron is derived from Arabic ' *zá-faran'* which means be yellow . [2]
At the ancient times Saffron has been also used as a drug to treat various human health conditions such as coughs, stomach disorders, colic, insomnia, chronic uterine haemorrhage, femine disorder, scarlet fever, smallpox, colds, asthma and cardiovascular disorders. [3]
Saffron is a triploid sterile species with 3n=24 ,x=8 that is vegetatively propagation means of corms.
A corm survives for only one season, reproducing via division into "cormlets" that eventually give rise to new plants.

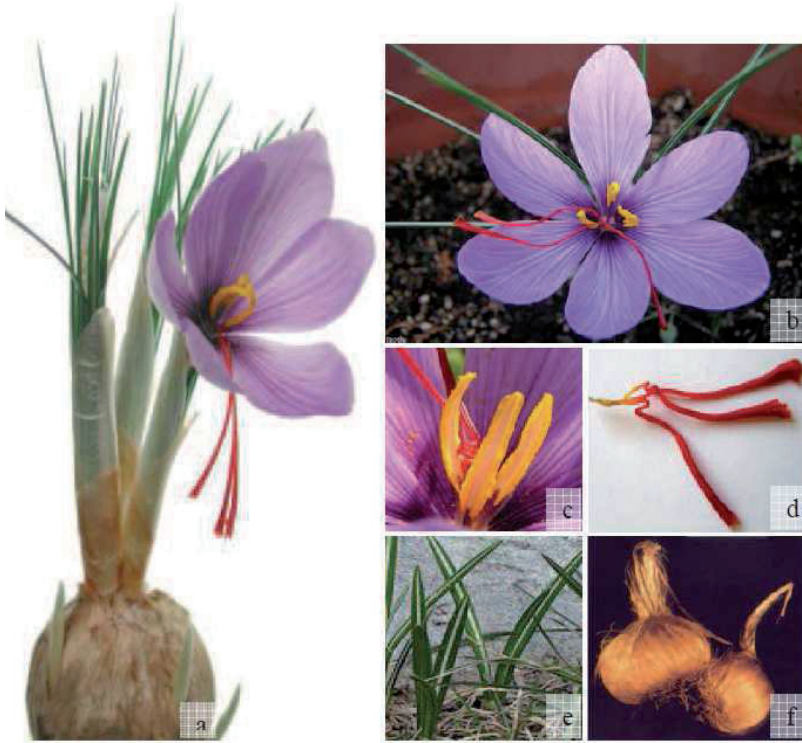

(Fig1.1) Saffron morphology

a. General appearance of saffron.,b. Purple flowers have three violet sepals and three petals similar together., c. Three distinct yellow stamen. ,d. orange red three stigma., e. Grass-like and dark green leaves. ,f .Saffron corm,(available at :www.webshot.com)

The growth of Saffron is under special climates ,although it needs hot temperature and dry climates its vegetative growth needs cold weather.

The saffron plant is characterized by biological cycle with a long pause in the summer and an active growth period in the autumn. Corms are dormant (leafless) in summer although flower differentiation occurs at this time.

Apex differentiation progresses further in May to July, during which time intensive laying down of generative organs occurs. The flowering of the maternal corm coincides with the formation of daughter corms during November; pale lilac flowers appear in autumn and may occur before, at the same time as, or after, leaf appearance.[4].

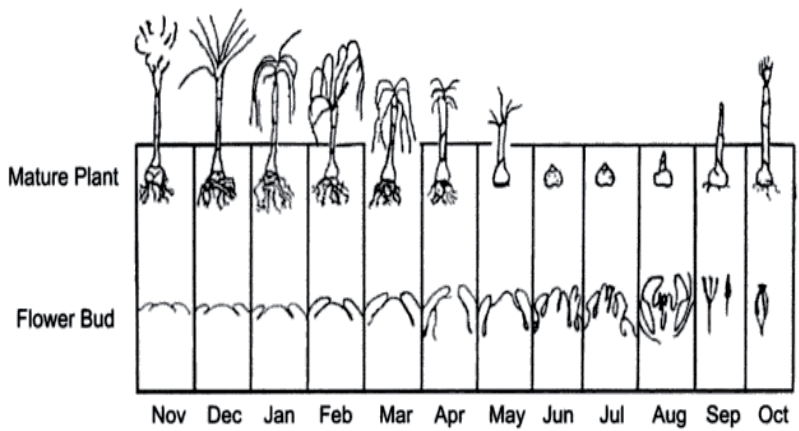

(Fig 1.2) shows the period of Saffron growth in one year. The appearance of flowers occure during Autumn (Nov-Dec), Corms are dormant(leafless) during Summer (Jun-Aug). Leafs occur at the end of Summer (Sep).[1]

---

[1] (. saffron production and processing by M .Kafi, page.44 )

Despite wide use of this unique spice, it is highly priced due to high demand and low supply. The main reason for its great cost is that saffron is still propagated by human help.

Each saffron flower has three stigmata and one stigma weights about 2mg. It takes 150,000-200,000 flowers and over 400 h of hand labor to produce 1kg saffron stigma [5].

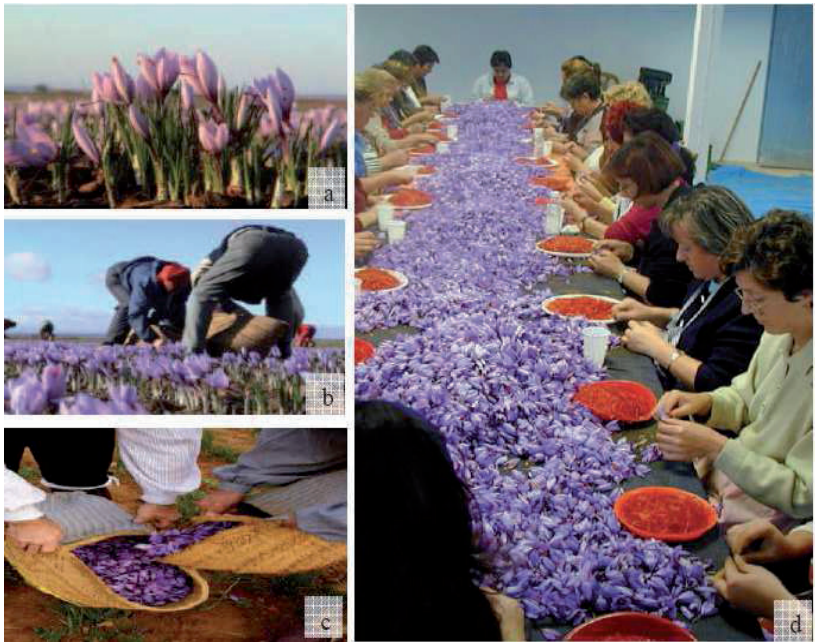

(Fig1.3) The cultivation process of saffron

a ,b and c. Flower harvesting, d. Separation of the stigmas

(saffron Spain webpage at :www.Saffron-Spain.com/ingles/azafran.html)

Saffron is currently being cultivated more or less intensely in Iran, India ,Greece , Morocco,Spain,Italy,France,Switzerlan,Israel,Pakestan,Azarbaijan,China,Egypt , Japan and recently in Australia.

The world's total annual Saffron production is estimated as 205 tons per year. Iran is said to produce 80 percent of total :i.e.,160 tons ,and Khorasan province alone 137 tons of total .The Keshmir region in India produce between 8 to 10 tons ,Greek production is 4 to 6 tons per year . Morraco produces between 0.8 and 1 ton, and the production of Spain is about 0.3-0.5tons.

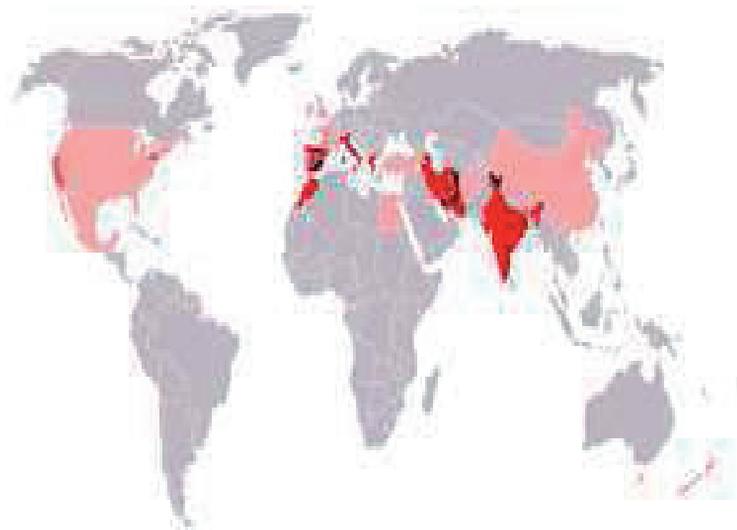

(Fig1.4)

Distribution and major producer of saffron

    ■ Major growing region
    ■ Improvable growing region
    ■ Minor growing region

(adopted from :http;//Wikipedia.org/wiki/saffron)

## 2) Morphology

Each Saffron plant has 4-8 green leaves with distinct white median stripe. The leaves are about 10-23 cm in size. The bisexual ,sterile flowers are usually one, sometimes three in number. Each flower has three violet sepals and three petals similar together.. The anthers are yellow and 8-12 mm in size.

The filament is yellow or pale orange and 2-5 mm in size. The three stamens are located opposite to outer tepal whorle.

The gynoecium is composed of three carpels, showing a single three-branched style and inferior ovary.

Saffron is not able to set viable seed. There for ,corms are indispensable for propagation of saffron. These corms are three to five cm in diameter and covered by tunics. Corms consist of nodes and are internally made up of starch-containing parenchyma cells.

Saffron corms produce both fibrous roots and contractile roots. The fibrous roots emerge from a single ring at the base of corms. These roots are straight and thin, 1 mm In size.

The contractile roots has the appearance of a tuber organ , very large and whitish. This type of roots enable corms to dig into the ground ,so corms rest in optimum depth and position in soil. [6]

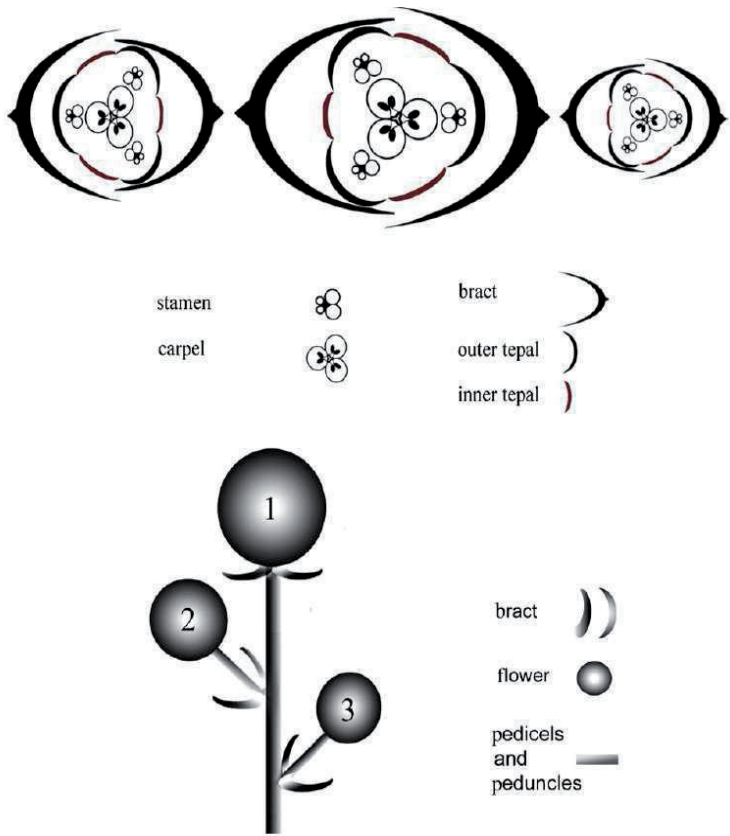

(fig 1.5) diagrams in polar and side views of a three-flowered Crocus inflorescence. [7]

## 3 ) saffron medical use

Herbal medicine can be a valuable source of assistance for traditional medicine. There are a number of herbs that can be used in conjunction with modern medicine like *Crocus sativus* L. The use of saffron as a drug goes back to the ancient times. It has used to treat various human health conditions such as coughs, stomach disorders, colic, insomnia, chronic uterine haemorrhage , femine disorder ,scarlet fever, smallpox, colds, asthma , and it has also used as a protective agent against chromosomal damage , high blood pressure.[8].

Modern medicine has also discovered saffron as having anticarcinogenic (cancer-suppressing),anti-mutagenic (mutation-preventing),anti-inflammation of internal organ like liver , immune-modulating and antioxidant-like properties.

Saffron has been reported to inhibit an increase in triglycerides, total LDL , cholesterol and keep cholesterol levels healthy, therefore is helpful in maintaining healthy arteries and blood vessels ,so it is shown that saffron helps reduce the risk of heart diseases by strengthening the blood circulatory system and promote healthy heart and prevent different cardiac problems.

Recent studies have shown the beneficial effects of saffron in depression, premenstrual syndrome (PMS), Parkinson and Alzheimer's Disease. Saffron extraction,.[9].

Today, based on growing and effective application of saffron in medical fields and alternative medicine , has attracted the attention of many researchers .

Saffron may substitute chemical medicines. Some medical properties of saffron that attract modern medicine are as follows:

 helps digestion, strengthens the stomach and is anti-tympanites , activates the sexual desire, is analgesic, especially for colicky pains gingivitisis euphoriant and alleviates neuralgia, is a tranquilizer, improve concentration, reacts against spasm, cures iron deficiency (anemia) in girls, macula lutea and retinopathy ischemic caused by old age. Cures rheumatism and bruises when used externally, cures amebic dysentery, measles, splenomegaly and urogenital infections. also saffron can increase the bioavailability and enhance absorption of other drugs.[10]

## 3-1 ) Toxicity and adverse effects of saffron

Findings of *in-vivo* studies have revealed that saffron has negligible toxicity. Oral $LD_{50}$ of saffron decoction in mouse has been reported to be 20.7 g/kg. Higher doses could be lethal due to the toxic effects on central nervous system and kidneys. Oral administration of saffron extract at doses between 0.1-5.0 g/kg has been reported to be non-toxic in mouse model. Clinical data on the toxicity and safety of saffron have been inconsistent. Daily consumption of saffron up to 1.5 g/day has not been found to be associated with any adverse effect. However, doses higher than 5 g are toxic, and at 20 g are lethal. Saffron doses over 10 g have been used for abortion with high risk of maternal death. At this latter dose, saffron can induce vomiting, uterus bleeding, hematuria, gastrointestinal bleeding, and vertigo.[11]

The most frequent adverse effects of saffron mentioned in studied books were headache, nausea, head fullness, dizziness, hypomania, and appetite suppression .yellowing is another side effect reported for saffron. Modern scientific studies have also implied that colored constituents of saffron may accumulate in sclera, skin, or mucosa, thus mimicking icteric complaints [12].

# Chapter 2

# Tissue Culture & Molecular Research

## 2 . Tissue culture & Molecular research
### 2.1. *Tissue culturing*

Tissue culture is the growth of tissue or cells separate from organism ,it is typically facilitated via use of a liquid or solid growth agar medium.

In vitro propagation of pathogen –free reproduction organs has been reported in some geophytes including saffron.

Schenk and Hildebrandt (1972) reported the importance of medium composition and techniques for induction and growth of monocotyledonous and dicotyledonous plants in cell culture.

As we know saffron is a monocotyledon member of the larg family Iridaceae. Within the last few decades, an increasing number of bulbous and cormous monocotyledons have been successfully cultured. Tissue culture technology was greatly influenced by the demand of rapid multiplication and clonal propagation of slow-growing monocots. Several economically important monocot species constituting nutritional, medicinal or ornamental groups of plants were used for *in vitro* clonal propagation .[13]. and production of secondary metabolites.[14].

In this study, Flower samples were conducted from Ghaen district of Southern Khorasan province during early Autumn of 2012.

As we need the internal tissues of the flower buds, like perianth ,style and ovary ,we separate corms and saffron flower buds.

Floral buds thoroughly washed in running tap water(30 min) and then sterilized with disinfection fluid containing 0.5% benzylkonium chloride ,(15 min.),70% ethanol (2min.), sodium hypochlorite 1% supplemented with few drops of tween 80 (20 min.) and finally rinsed 3 times with sterile distilled water .

Perianth organs were as specially separate from the sterilized buds and cultured on petri dishes containing MS [2], supplemented with 10mg/l NAA and 10 mg/l BAP for induction of SLSs (Stigma Liked Structure). All cultures were kept under 22±2 °C temperatures in darkness ,calli were subcultures every 28 days and collected from three development stages , freezed-dried and then stored at -80 °C for further uses.[15].

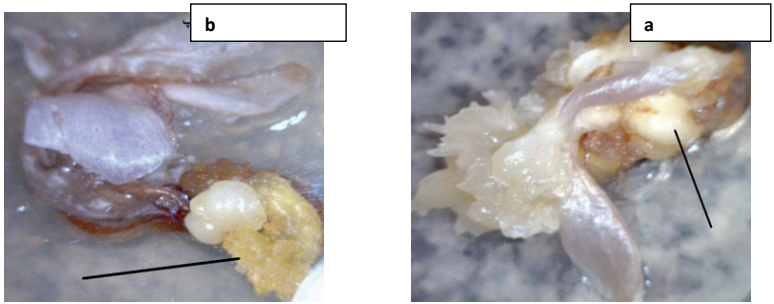

(Fig 2.1) Colorless calli at stage I (SI) without SLSs, **a**, inflation and transformation of perianth explants after 2 months

**b**, white perianth callis become yellow after 3 months.

---

[2] (Murashing Skoog,1962)

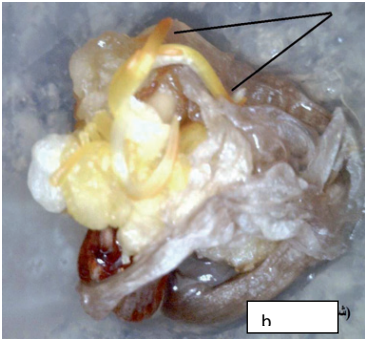 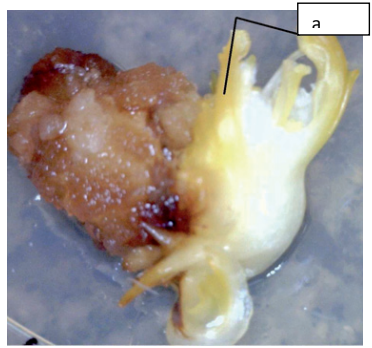

(Fig 2.2). Pale yellow calli at stage II (SII) on which SLSs were initiate, **a, b**. perianth SLSs after 4 months.

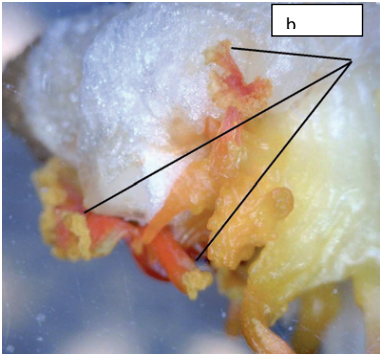 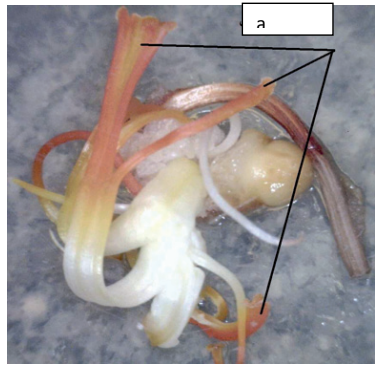

(Fig2.3) Fully developed red SLSs on calli at stage III (S III). **a,b** perianth SLSs after 6 months , separated and stored at -70˚C ,and are used for molecular investigation in this project.

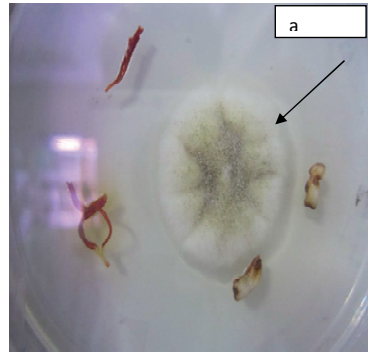 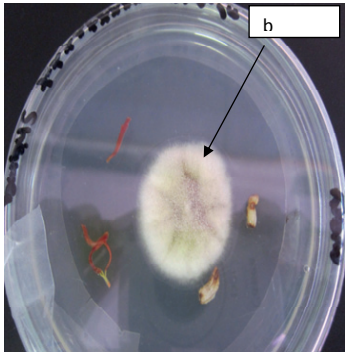

(Fig 2.4) a,b :white spots, are samples of bacterial contamination in petri dishes media ,these saffron explants will never growth .*the fungal contamination are green and black.*

The One-way ANOVA, Tukey test indicates that, At <0.01 level the callus number has significant difference, and between second and third month the number of contractile root has significant difference at <0.05 level.

the maximum number of Perianth callus are in fifth and half month,(80%) a bit higher that number of contractile roots (79%),the emergence of SLSs start in third month and achieve its maximum amount in fifth month(32%).(Diagram 2.1). [3]

---

[3] . ** all of the explants(Style,Stigma,Ovary and perianth) were cultured and sub-cultured , their morphological changes were measured and compared carefully with ANOVA and tukey test. But here I just bring the perianth's data because of the title of the book.

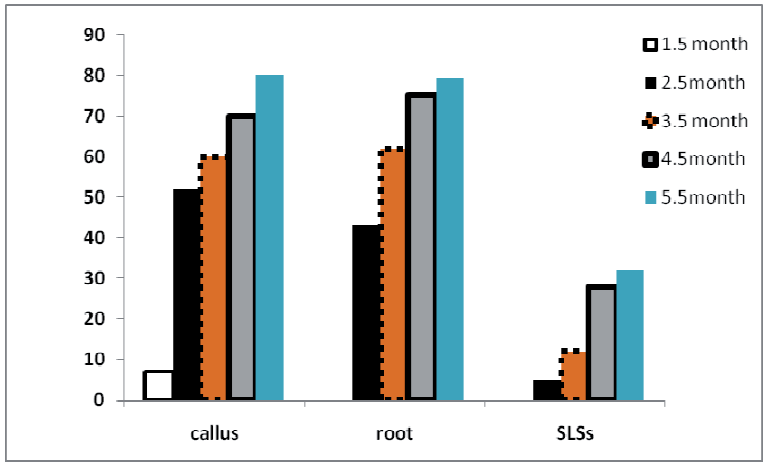

(Diagram2.1) investigation of morphological changes : number of callus ,contractile root and SLSs ,during 6 month of tissue culturing in perianth explant of Ghaen field.

## 2.2 Implication of Carotenoid Biosynthesis Genes during stigma development

Carotenoids are terpenoid compounds that are ubiquitous in nature. In all photosynthesis organism, they carry out essential function in light harvesting systems and photosynthetic reaction centers. In higher plants ,carotenoids play additional roles in providing distinc yellow,orange and red colors to certains organs,such as flowers and fruits. In those tissues ,unique carotenoids synthesized as secondary metabolites accumulate to high concentration and are stored within the chromoplast. All isoprenoids ,including carotenoids ,are derived from the ubiquitous C5 building blocks isopentenyl diphosphate and dimethylallyl diphosphate .these precursors can be synthesis through two different routes : the classic mevalonate pathway in cytoplasm or alternative non-mevalonate pathway in plastids.[16],[17].(Fig2.5).

The plastid pathway ,now known as the 2-C-methyl-D-erithritol-4-phospate pathway.

The first commiting step of the carotenoid biosynthesis is a head to head coupling of two geranylgeranylphosphate to yield colorless phytoene by phytoene synthesis (PSY).

Subsequently ,four addition double bonds are introduced by phytoen desaturase (PDS) and zeaxanthin desaturase (ZDS) producing the colored carotenes phytofluene,neurosporene and lycopene.

The cyclization of lycopene creates a series of carotenes that have one or two rings of either the β- or ε type. Lycopen β-cyclase (LYCb) catalyzes a two-step reaction that leads to β-carotene (two β-rings),whereas lycopen ε-cyclase creats one ε-ring to produce δ-caroten .

The resulting compounds are further modified by the introduction of hydroxyl groups onto the ionone rings.one ε-ring and one β-ring hydroxylation of α-carotene yield lutein in reactions catalyzed by the ε-hydroxylase, a member of the cytochrome P450 and the β-hydroxylase (BCH). Two β-ring hydroxylations of β-carotene yield zeaxanthin in a reaction catalyzed by BCH.

Two enzymes involved in saffron is caused by apocarotenoids,specially crocetin ester with gentobiose. These saffron apocarotenoids are formed by zeaxanthin cleavage,followed by specific glucosylation steps. (Fig 2.6)

Crocus zeaxanthin 7,8(7′,8′)-cleavage dioxygenase gene (CsZCD) which codes for chromoplast enzyme that initiates the biogenesis of crocetin glycosides and picocrocin is the gene that we study the existence and expression of it, in SLSs perianth callus of saffron. (Fig 2.7).

In C.sativus stigmas, the final step involves glucosylation of the generated zeaxanthin cleavage products by glucosyltransferase 2 enzyme which is coded by (CsUGT2) gene in chromoplast of stigmas and then sequestered into the central vacuole of the fully development stigmas.

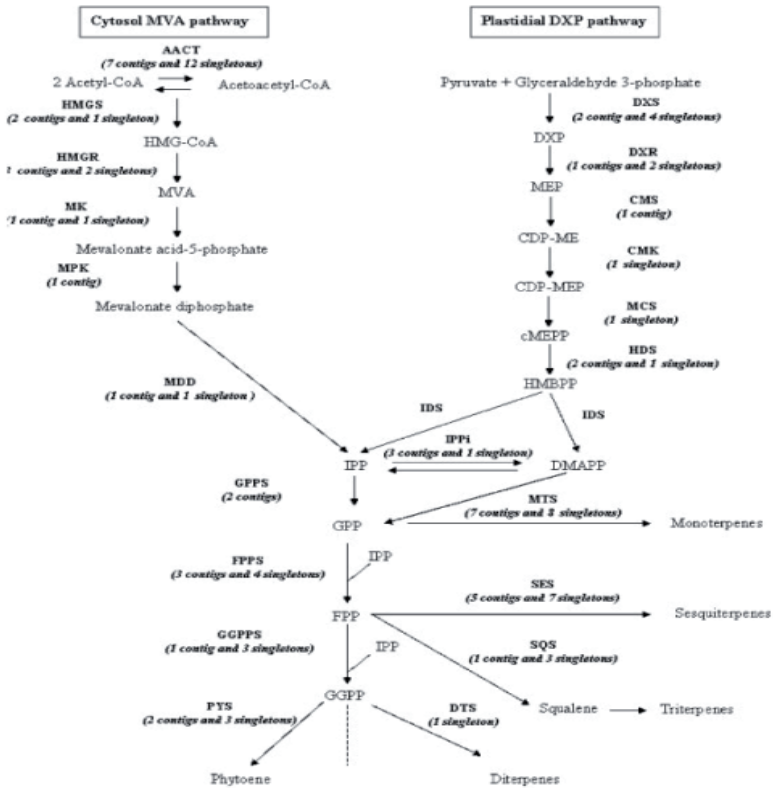

(Fig 2.5) mevalonic and non-mevalonic pathways that occure in cytosol and plastids.[18].

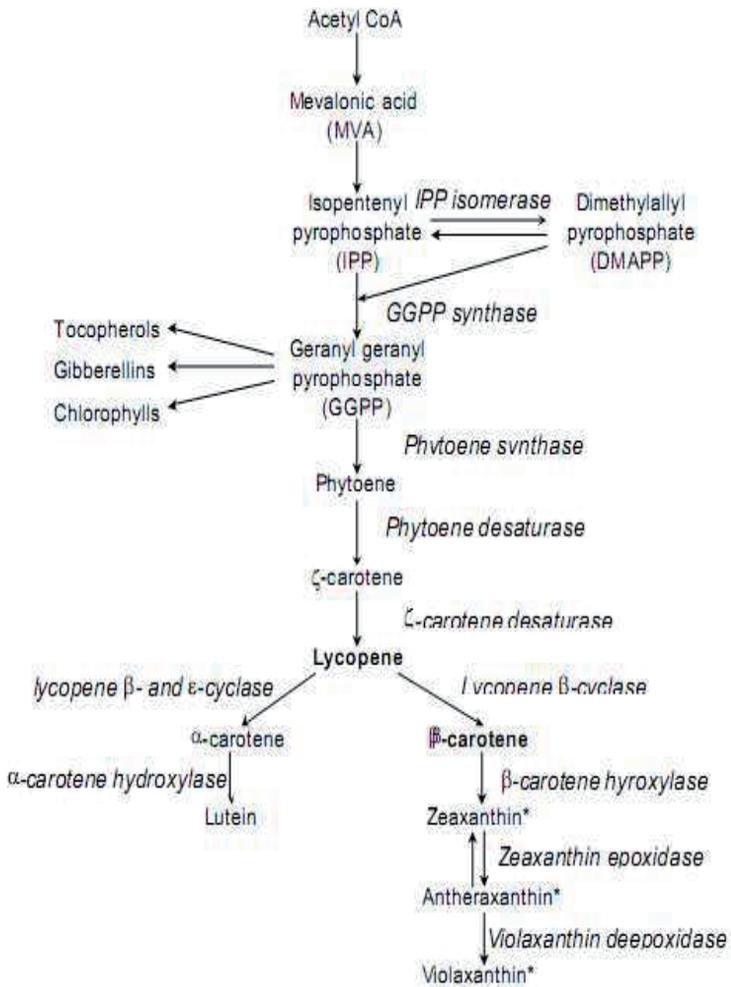

(Fig 2.6) Biosynthetic Mevalonate pathway [4].

---
[4]. Naik et al., 2003

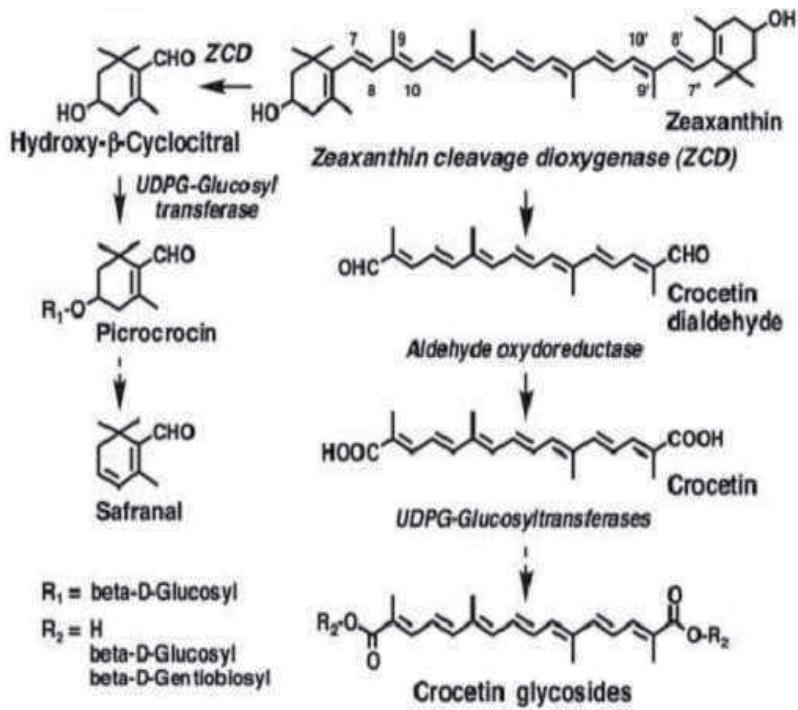

(Fig 2.7) the biosynthesis pathway of second methabolites and ZCD function.[5]

The components of the spice "saffron" are localized in the red stigmatic lobes of C. sativus flower and these are responsible for its distinct color, flavor and smell [19].

For color the principal pigment is crocin, for smell the main component is safranal and for the special bitter flavor the main compound is the glycoside picrocrocin.[20].( Fig 2.8).

---

[5]. Bouvier et al., 2003

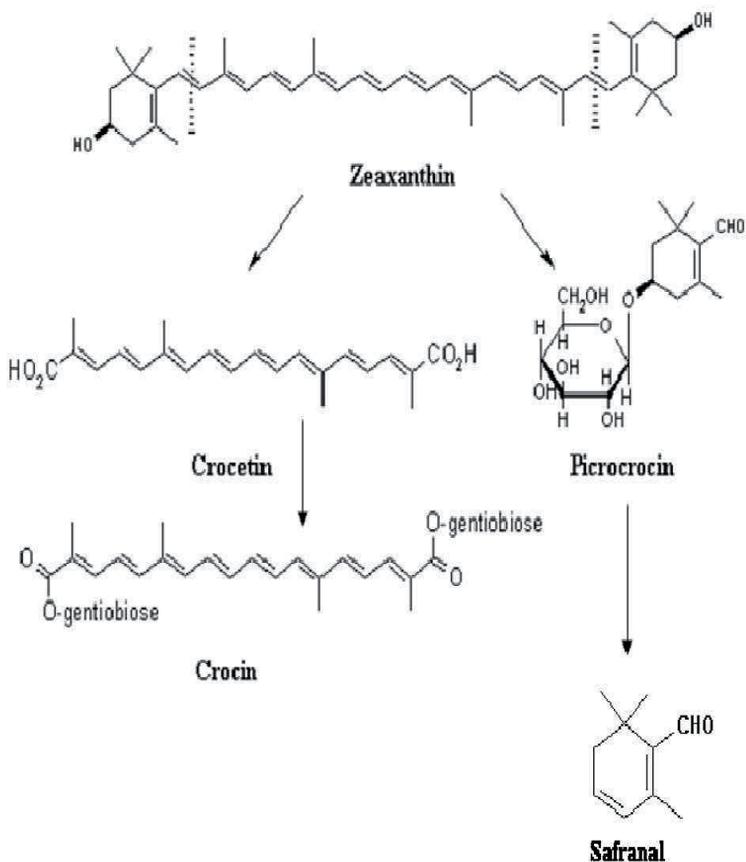

(Fig 2.8). The biogenesis of the crocin and picrocrocin which are derived from the bio-oxidative cleavage of Zeaxanthin at points 7, 8-(7', 8'). Safranal is produces by de-glycosylation of picocrocin. [21].

The list of the key genes of mevalonic acid pathway are mentioned below :

1- GGPP synthase (*Ggps*)
2-Phytoene synthase *(PSY*)
3-Phytoene desaturase (*PDS*)
4- z-carotene desaturase (*ZDS*)
5-Lycopene b-cyclase (*LCY*)
6-carotenoid cleavage 9, 10 (9´,10´) dioxygenase gene (*CCD*)
7- β-carotene hydroxylase (*BCH*)
8- zeaxanthin cleavage 7,8 (7´,8´) dioxygenese (*ZCD*)
9- glycosyltransferase (*UGT*)

## 2.3 RNA extraction

RNA extraction is the first step of *In vitro* expression investigation of apocarotenoid genes in this process.
Freeze dried perianth callus from stage III separated and total RNA was extracted using RNx method (without kit) following the manufacturer's protocol. (Table 2.1). Quality of the extracted RNAs was checked by measuring the absorbance at 260 and 280 nm by spectrophotometer and RNAs ratio of OD260/OD280 ranging from 1.2 to 1.5 were used for cDNA synthesis .

(table 2.1) RNA extraction protocol by RNX<sup>TM</sup> –PLUS.

- Keep 0.3 -0.5 g the tissue frozen in liquid Nitrogen until the homogenization procedure is ready to be performed.
- Add 500-700 ml cold RNX<sup>TM</sup> –PLUS to 1.5 ml tube containing homogenized sample.
- Vortex 5-10 sec. and incubate at room tempreture for 5 min.
- Add 200 μl of chloroform.
- Mix well for 15 sec. by shaking .
- Incubate on ice for 5 min.
- Centrifuge at 12000 rpm at 4˚C for 20 min.
- Transfer the aqueos phase to new RNase-free 1.5 ml tube.
- Add equal volume of Isopropanol to the tube.
- Gently mix and incubate on ice for 15 min.
- Centrifuge the mixture at 12000 rpm at 4˚C for 15 min.
- Discard the supernatant and add 1 ml of 75% Ethanol.
- Centrifuge 7500 rpm at 4˚C for 8 min.
- Discard the supernatant and let the pellet to dry at room temperature for 20 min.(do not let dry completely).
- Dissolve pellet in 30μl of RNase free water or DEPC treated water ,to help dissolving ,place the tube in 55-60˚ C water bath for 10 min.
- Glassware should be treated before use to ensure that is RNase –free. Glassware use for RNA work should be cleaned with a detergent,thoroughly rinsed,and oven baked at 240 ˚C for at least 4 hours (over night),before use.  Alternatively ,glass wear can be treated with DEPC(diethyl pyrocarbonate). Fil glass wear with 0.1% DEPC (0.1% in water),allow to stand over night (12 hours) at 37˚ C,then outoclave or heat to 100˚C for 15 minutes to eliminate residual DEPC.

| Tissue | OD 260/280 | OD 260/230 | Conc. |
|---|---|---|---|
| Perianth callus (SLSs) | 1.95 | 1.08 | 2134 |

(Table 2.2) Quality of the extracted RNAs was checked by measuring the absorbance at 260 and 280 nm by Nanodrop(based on spectrophotometery)

Although it is possible to estimate the concentration of solution of nucleic acid and oligonucleotides by measuring their absorption at a single wavelength (260 nm) this is not good practice.

The absorbance of the sample should be measured at several wavelengths since the ratio of absorbance at 260 nm to absorbance at other wavelength is a good indicator of purity of preparation.

Significant absorption at 230 nm indicate contamination by phenolate ion, thiocynates and other organic compound [6], whereas absorption at higher wavelength (230 nm and higher) is usually caused by light scattering and indicates the presence of particulate matter. Absorbance at 280 nm indicates the presence of protein ,because amino acids absorb strongly at 280 nm.

Pure preparation of RNA have OD 260: OD 280 value 1.8 and 2.0.

The Ribosomal RNA profile were visualized with ethidium bromide staining following agarose gel 1% electrophoresis. (Fig 2.9).

---

[6] (Stulnig & Ambeger 1994)

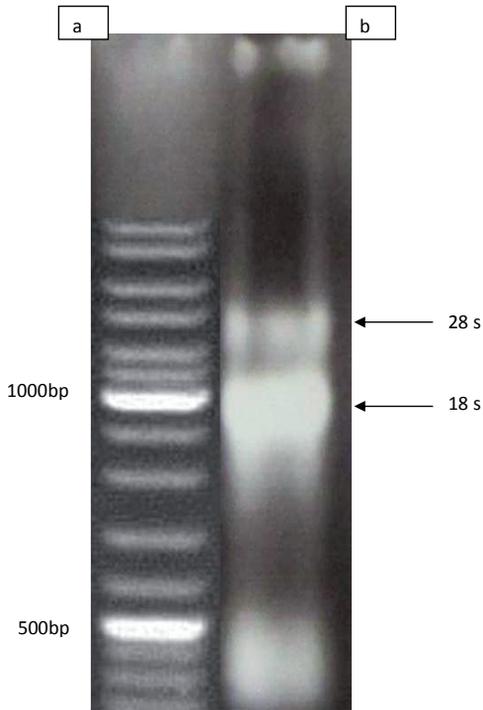

(Fig 2.9) a. ladder with obvious base pare rates(1000bp, 500 bp) , b. The bands (28 s , 18 s) show the existence of RNA on agarose gel .

We can store this RNA in 1.5 ml tube at -70˚ C , for synthesizing cDNA.

## 2.4 ) cDNA preparation:

cDNA [7], is DNA synthesized from a messenger RNA(mRNA) template in a reaction catalysed by enzymes reverse transcriptase and DNA polymerase . (Fig 2.10).

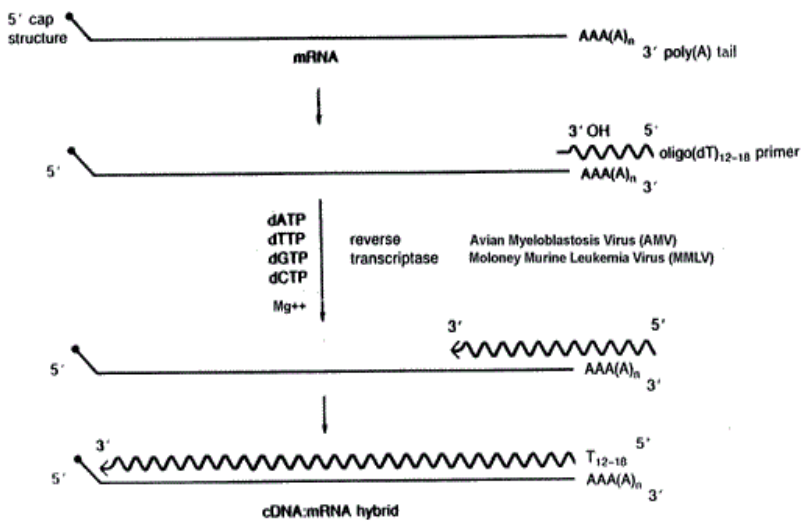

**(Fig 2.10)** The reverse transcriptase enzyme is an RNA-dependent DNA polymerase isolated from a retrovirus (AMV or MMLV). As with other polymerases a short double-stranded sequence is needed at the 3' end of the mRNA which acts as a start point for the polymerase. This is provided by the poly(A) tail found at the 3' end of most eukaryotic mRNAs to which a short complementary synthetic oligonucleotide (oligo dT primer) is hybridized (polyT-polyA hybrid). Together with all 4 deoxynucleotide triphosphates, magnesium ions and at neutral pH, the reverse transcriptase synthesises a complementary DNA on the mRNA template.
(from Molecular Genetics **II** – Genetic Engineering course**.**

---

[7] . Complementary DNA

For each samples 8μl of total RNA was used for first-strand cDNA synthesis as describe by the protocol. (Table 2.3).

**Master 1** :

| RNA | 8 μl |
|---|---|
| Oligo dT | 1μl |
| DEPC water | Minimizing the volume |
| Total volume | 13.5 μl |

**Master 2** :

| dNTP | 1 μl |
|---|---|
| Buffer | 2 μl |
| RT enzyme | 1 μl |
| RNase inhibitor | 0.5 μl |
| DEPC water | 2 μl |
| Total volume | 6.5 μl |

(table 2.3) cDNA synthesis protocol. master 1 should store 10 min. in Hot palte. Then we should add master 2 to the master 1.[8]

The synthesized cDNA should be stored at -20 ˚C for gene expression study.

---

[8] . ** these volumes are used for *Crocus sativus* L. so for other plant's species these are variable.

## 2.5 RT- PCR

PCR [9] provides an extremely sensitive means of amplifying small quantities of DNA. The development of this technique resulted in an explosion of new techniques in molecular biology[10] . The technique was made possible by the discovery of Taq polymerase, the DNA polymerase that is used by the bacterium *Thermus auquaticus* that was discovered in hot springs . This DNA polymerase is stable at the high temperatures need to perform the amplification, whereas other DNA polymerases become denatured.

Since this technique involves amplification of DNA, the most obvious application of the method is in the detection of minuscule amounts of specific DNAs, This important in the detection of low level bacterial infections or rapid changes in transcription at the single cell level, as well as the detection of a specific individual's DNA in forensic science (like in the O.J. trial). It can also be used in DNA sequencing, screening for genetic disorders, site specific mutation of DNA, or cloning or sub-cloning of cDNAs.

The method relies on thermal cycling ,consisting of cycle of repeated heating and cooling of reaction for DNA melting and enzymatic replication of the DNA. Primers [11] containing sequences complementary to the target region along with a DNA polymerase. Two primers that are complementary to the 3 ends of each sense and anti – sense strand of DNA target are used.

---

[9] The Polymerase Chain Reaction

[10] The Nobel prize for Kary Mullins in 1993

[11] Short DNA fragments ,

A primer and dNTPs are added along with a DNA template and the DNA taq polymerase.

PCR based on three major levels :

The original template is melted at 94°C, the primers anneal at 45-55°C, and the polymerase makes two new strands at 72°C , doubling the amount of DNA present. This provides 2 new templates for the next cycle. The DNA is again melted, primers anneal, and the Taq makes 4 new strands. (Fig 2.11)

A basic PCR set up requires several other components and reagents that includes:

**dNTP[12] : the building – blocks from which the DNA polymerase synthesized a new DNA strand.

** Buffer solution: providing a suitable chemical environment for optimum activity and stability of the DNA polymerase.

**Mangnesium or Manganese ions ,generally $Mg^{2+}$ is used, but $Mn^{2+}$ can be utilized for PCR-mediated DNA mutagenic , as higher $Mn^{2+}$ concentration increases the error rate during DNA synthesis.

For each samples, 5μl of cDNA was used for PCR reaction, as describe by the protocol.(Table 2.4).

---

[12] . Deoxynucleoside triphosphates, nucleotides containing triphosphate groups

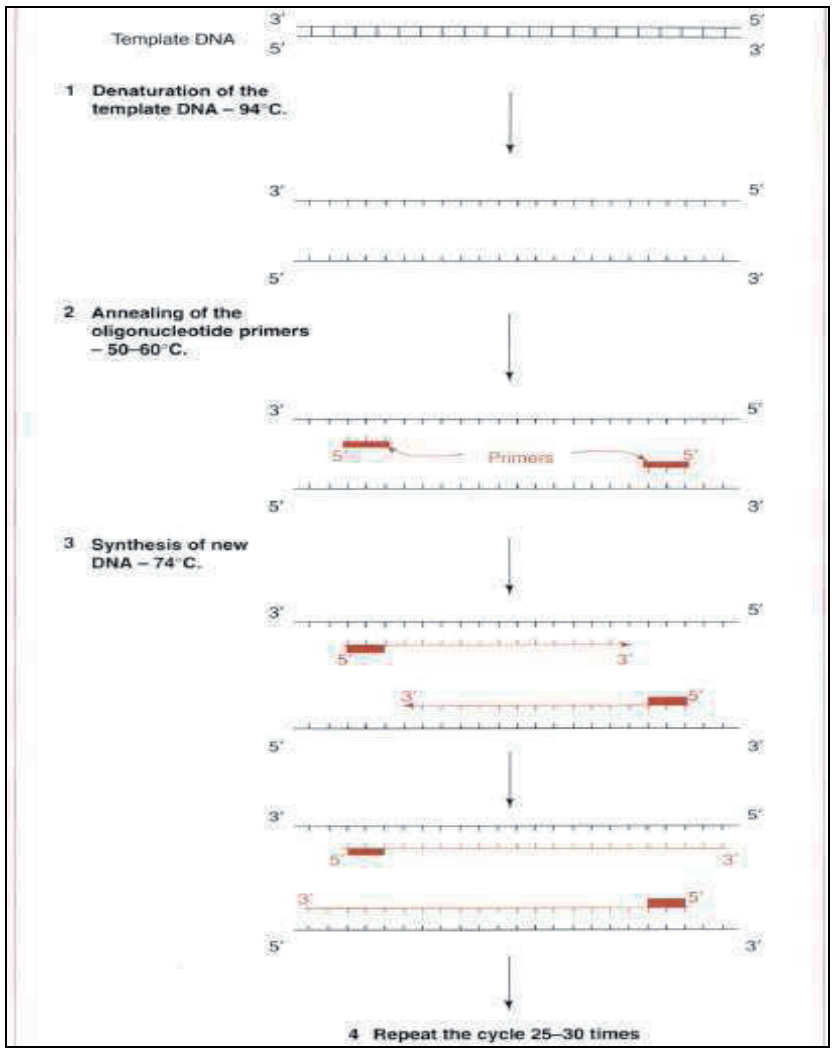

(Fig 2.11) polymerase reaction steps ,PCR based on three major steps : a, denaturation of template at 94 ˚C, b, annealing of oligonucleotides primers at 50-60 ˚C, c, synthesis of mew DNA by DNA polymerase enzyme at 74 ˚C.(the picture is from T.A.Brown book).

| Buffer 10 X | 2.5 µl |
|---|---|
| dNTP | 0.5 µl |
| Mgcl2 | 0.75 µl |
| Primer forward | 0.5 µl |
| Primer reverse | 0.5 µl |
| Taq pol enzyme | 0.25 µl |
| cDNA | 5 µl |
| DEPC water | 15 µl |
| Total volume | 25 µl |

(table 2.4) PCR synthesis protocol. The Taq polymerase enzyme is so sensitive ,it is better to add this enzyme in the last step,and at 20˚ C.[13]

Amplification was carried out using gene-specific forward and reverse primers ,cDNA obtained as templates For amplification of *CsZCD* , as well as *CsTUB* as internal control according to the manufacturer's instructions. Gene specific primers were designed to flank introns. [22].[23].

Designed forward and reverse primers were:

5'-GTCTTCCCCGACATCCAGATC-3' and 5TCTCTATCGGGCTCACGTTGG-3' for *CsZCD* gene . (GenBank access No. AJ489276)

and 5'-ATGATTTCCAACTCGACCAGTGTC- 3' and
5'-ATACTCATCACCCTCGTCACC ATC-3' for *CsTUB* gene as a internal control.
(GenBank access No. AJ489275). [14]

---

[13] .** The volume of these reagents are used for *Crocus sativus* L. and are not suitable for other species.

[14] . (Bouvier et al., 2003; Castilla etal., 2005).from "invitro expression of apocarotenoid genes in *crocus sativus* L. paper by Monir.H.Namin.18 Sep.2009.

The length of the products were : 241 bp for *CsZCD* and 225 bp for *Cs TUB*.

according to the following conditions: 5 μl of cDNA was used. Initial denaturizing at 95°C for 5min followed by 35 cycles of amplification according to the subsequent scheme; denaturizing 1 min. at 94°C, annealing at 56.2°C for 30 s and extension at 72°C for 40 s. and final extension at 72°C for 10 min.
The experiments were repeated twice.
Subsequently 3 μl of the PCR products were used on 1% (w/v) agarose gels electrophoresis. The images of stained gels with ethidium bromide were scanned and captured by a gel-documentation System. (Fig 2.12 .a.& b.)

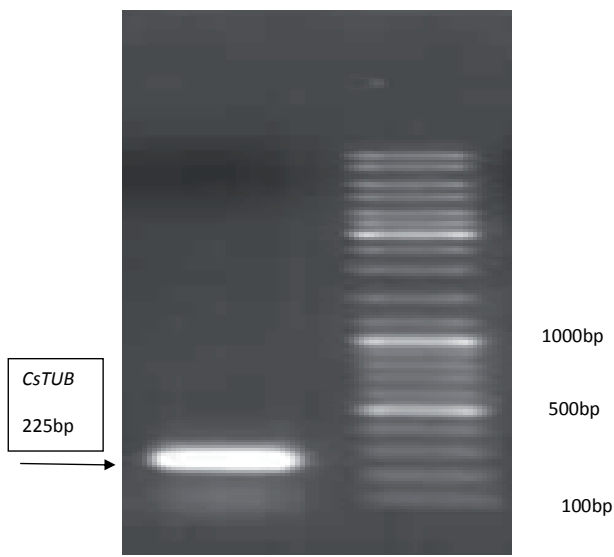

(Fig 2.12.a) expression level of *CsTUB* gene as internal control in fully developed red SLSs of *C.sativus*.

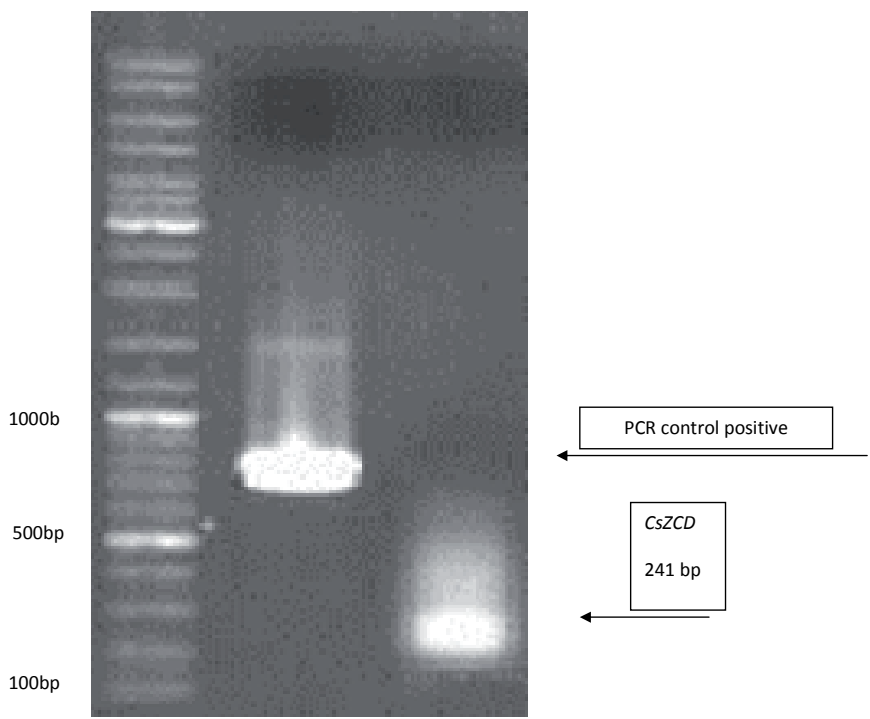

(Fig 2.12.b) expression level of *CsZCD* gene at 241 bp and pcr control positive in fully developed red SLSs of *C.sativus.*

This study confirmed that the ZCD gene is present in perianth callus, and is expressed in Stigma liked Structure (SLS)of callus. By using callus with mature and red SLS to evaluate the in vitro expression of csZCD which is involoved in apo-carotenogenesis of stigma in Mevalonic acid pathway (MVA).

# Chapter 3

# CONCLUSION & SUGGESTION

## 3- conclusion & suggestion
## 3-1 conclusion

In this study the *Crocus sativus* L. flower buds brought from Ghaen district of Southern Khorasan provience during early autumn.

We have separated organs and cultured them in MS medium, each month we sub-cultured them and measured their morphological features. We compared them with each other.

By tissue culturing we could produce saffron anytime we want although the climates or fields are not appropriate.

After six month we could have lots of SLSs from one explants but in nature we should wait at least one year ,it means that by using only one plants,it is possible to produce more than one thousand of same plants.

It was obvious that our production were without infection and there were suitable for molecular research.

This study confirmed that the ZCD gene is present in mature perianth callus at stage lll, and is expressed in Stigma liked Structure (SLS)of these callus. By using callus with mature and red SLS to evaluate the invitro expression of csZCD wich is involoved in apocarotenogenesis of stigma in Mevalonic acid pathwat(MVA). According to Bouvier et al. (2003), our results based on *CsZCD* gene expression showed that this gene to be identical to fully developed natural stigmas.

## 3- conclusion & suggestion
## 3-2 . suggestion

- Investigation the intensity of expression of *CsZCD* gene by Real Time – PCR method and compare with RT-PCR method.[15]

- The investigation of other MVA pathway genes includes: *CsPSY,CsPDS,CsZDS,CsCCD* and *CsUGT* in other explants callus like: Corm,Stigma,Ovary,Style,stamen,…[16]

- Investigation the effects of the stress on the *in vitro* gene expression in MVA[17] pathway.

- Investigation of in vitro starch degrading enzyme genes expression in amiloplast conversion to chromoplast by emergence of SLSs in explants.

---

[15] We are working on this way and the result will publish as soon as possible.(Authors)

[16] We have investigated the CsLYC,CsBCH genes on Style,ovary and perianth and the results will publish in our next books.(Authors).

[17] Mevalonic acid pathway

# Appendix A

| COMPONENTS | mg/L |
|---|---|
| **Microelements** | |
| $CoCl_2 \cdot 6H_2O$ | 0.025 |
| $CuSO_4 \cdot 5H_2O$ | 0.025 |
| FeNaEDTA | 36.70 |
| $H_3BO_3$ | 6.20 |
| KI | 0.83 |
| $MnSO_4 \cdot H_2O$ | 16.90 |
| $Na_2MoO_4 \cdot 2H_2O$ | 0.25 |
| $ZnSO_4 \cdot 7H_2O$ | 8.60 |
| **Macroelements** | |
| $CaCl_2$ | 332.02 |
| $KH_2PO_4$ | 170.00 |
| $KNO_3$ | 1900.00 |
| $MgSO_4$ | 180.54 |
| $NH_4NO_3$ | 1650.00 |
| **Vitamins** | |
| Glycine | 2.00 |
| Myo-inositol | 100.00 |
| Nicotinic acid | 0.50 |
| Pyridoxine HCl | 0.50 |
| Thiamine HCl | 0.10 |
| **TOTAL** | 4405.19 |

(Table A1) composition of MS basal medium(microelements,macro elements and vitamins)

# Appendix B

**Test of Homogeneity of Variances**

|  | Levene Statistic | df1 | df2 | Sig. |
|---|---|---|---|---|
| Callus numbers | 9.517 | 4 | 46 | .000 |
| Contractile roots | 7.016 | 4 | 46 | .000 |
| SLSs | 3.044 | 4 | 46 | .026 |

**ANOVA**

|  |  | Sum of Squares | df | Mean Square | F | Sig. |
|---|---|---|---|---|---|---|
| Callus number | Between Groups | 4514.453 | 4 | 1128.613 | 233.303 | .000 |
|  | Within Groups | 222.527 | 46 | 4.838 |  |  |
|  | Total | 4736.980 | 50 |  |  |  |
| Contractile .Roots | Between Groups | 3760.282 | 4 | 940.070 | 156.483 | .000 |
|  | Within Groups | 276.345 | 46 | 6.008 |  |  |
|  | Total | 4036.627 | 50 |  |  |  |
| SLSs | Between Groups | 850.627 | 4 | 212.657 | 212.657 | .000 |
|  | Within Groups | 46.000 | 46 | 1.000 |  |  |
|  | Total | 896.627 | 50 |  |  |  |

**Multiple Comparisons**

Tukey HSD

| Dependent Variable | (I) TIme | (J) Time | Mean Difference (I-J) | Std. Error | Sig. | 95% Confidence Interval | |
|---|---|---|---|---|---|---|---|
| | | | | | | Lower Bound | Upper Bound |
| Callus number | 1.5 month | 2.5 month | -5.955* | .961 | .000 | -8.68 | -3.23 |
| | | 3.5 month | -13.000* | .984 | .000 | -15.79 | -10.21 |
| | | 4.5 month | -18.900* | .984 | .000 | -21.69 | -16.11 |
| | | 5.5 month | -26.900* | .984 | .000 | -29.69 | -24.11 |
| | 2.5 month | 1.5 month | 5.955* | .961 | .000 | 3.23 | 8.68 |
| | | 3.5 month | -7.045* | .961 | .000 | -9.77 | -4.32 |
| | | 4.5 month | -12.945* | .961 | .000 | -15.67 | -10.22 |
| | | 5.5 month | -20.945* | .961 | .000 | -23.67 | -18.22 |
| | 3.5 month | 1.5 month | 13.000* | .984 | .000 | 10.21 | 15.79 |
| | | 2.5 month | 7.045* | .961 | .000 | 4.32 | 9.77 |
| | | 4.5momth | -5.900* | .984 | .000 | -8.69 | -3.11 |
| | | 5.5month | -13.900* | .984 | .000 | -16.69 | -11.11 |
| | 4.5 month | 1.5 month | 18.900* | .984 | .000 | 16.11 | 21.69 |
| | | 2.5 month | 12.945* | .961 | .000 | 10.22 | 15.67 |
| | | 3.5 month | 5.900* | .984 | .000 | 3.11 | 8.69 |
| | | 5.5 month | -8.000* | .984 | .000 | -10.79 | -5.21 |
| | 5.5 month | 1.5 month | 26.900* | .984 | .000 | 24.11 | 29.69 |

|   |   |   | 2.5 month | 20.945* | .961 | .000 | 18.22 | 23.67 |
|---|---|---|---|---|---|---|---|---|
|   |   |   | 3.5 month | 13.900* | .984 | .000 | 11.11 | 16.69 |
|   |   |   | 4.5 month | 8.000* | .984 | .000 | 5.21 | 10.79 |
|   |   |   | 2.5 month | -8.364* | 1.071 | .000 | -11.40 | -5.32 |
|   |   | 1.5 month | 3.5 month | -11.900* | 1.096 | .000 | -15.01 | -8.79 |
|   |   |   | 4.5 month | -19.000* | 1.096 | .000 | -22.11 | -15.89 |
|   |   |   | 5.5 month | -25.100* | 1.096 | .000 | -28.21 | -21.99 |
|   |   |   | 1.5 month | 8.364* | 1.071 | .000 | 5.32 | 11.40 |
|   |   | 2.5 month | 3.5 month | -3.536* | 1.071 | .015 | -6.58 | -.50 |
|   |   |   | 4.5 month | -10.636* | 1.071 | .000 | -13.68 | -7.60 |
|   |   |   | 5.5 month | -16.736* | 1.071 | .000 | -19.78 | -13.70 |
|   |   |   | 1.5 month | 11.900* | 1.096 | .000 | 8.79 | 15.01 |
| roots |   | 3.5 month | 2.5 month | 3.536* | 1.071 | .015 | .50 | 6.58 |
|   |   |   | 4.5 month | -7.100* | 1.096 | .000 | -10.21 | -3.99 |
|   |   |   | 5.5 month | -13.200* | 1.096 | .000 | -16.31 | -10.09 |
|   |   |   | 1.5 month | 19.000* | 1.096 | .000 | 15.89 | 22.11 |
|   |   | 4.5 month | 2.5 month | 10.636* | 1.071 | .000 | 7.60 | 13.68 |
|   |   |   | 3.5 month | 7.100* | 1.096 | .000 | 3.99 | 10.21 |
|   |   |   | 5.5 month | -6.100* | 1.096 | .000 | -9.21 | -2.99 |
|   |   |   | 1.5 month | 25.100* | 1.096 | .000 | 21.99 | 28.21 |
|   |   | 5.5 month | 2.5 month | 16.736* | 1.071 | .000 | 13.70 | 19.78 |
|   |   |   | 3.5 month | 13.200* | 1.096 | .000 | 10.09 | 16.31 |
|   |   |   | 4.5 month | 6.100* | 1.096 | .000 | 2.99 | 9.21 |
| SLSs |   | 1.5 month | 2.5 month | -1.800* | .437 | .001 | -3.04 | -.56 |

|  |  |  | | | | |
|---|---|---|---|---|---|---|
| | | 3.5 month | -2.400* | .447 | .000 | -3.67 | -1.13 |
| | | 4.5 month | -8.400* | .447 | .000 | -9.67 | -7.13 |
| | | 5.5 month | -10.600* | .447 | .000 | -11.87 | -9.33 |
| | 2.5 month | 1.5 month | 1.800* | .437 | .001 | .56 | 3.04 |
| | | 3.5 month | -.600 | .437 | .648 | -1.84 | .64 |
| | | 4.5 month | -6.600* | .437 | .000 | -7.84 | -5.36 |
| | | 5.5 month | -8.800* | .437 | .000 | -10.04 | -7.56 |
| | 3.5 month | 1.5 month | 2.400* | .447 | .000 | 1.13 | 3.67 |
| | | 2.5 month | .600 | .437 | .648 | -.64 | 1.84 |
| | | 4.5 month | -6.000* | .447 | .000 | -7.27 | -4.73 |
| | | 5.5 month | -8.200* | .447 | .000 | -9.47 | -6.93 |
| | 4.5 month | 1.5 month | 8.400* | .447 | .000 | 7.13 | 9.67 |
| | | 2.5 month | 6.600* | .437 | .000 | 5.36 | 7.84 |
| | | 3.5 month | 6.000* | .447 | .000 | 4.73 | 7.27 |
| | | 5.5 month | -2.200* | .447 | .000 | -3.47 | -.93 |
| | 5.5 month | 1.5 month | 10.600* | .447 | .000 | 9.33 | 11.87 |
| | | 2.5 month | 8.800* | .437 | .000 | 7.56 | 10.04 |
| | | 3.5 month | 8.200* | .447 | .000 | 6.93 | 9.47 |
| | | 4.5 month | 2.200* | .447 | .000 | .93 | 3.47 |

(table B.1) minimum, maximum, means and standard deviation value of the number of callus ,contractile roots and SLSs of perianth explants with one-way ANOVA,tukey test.